\documentclass[%
 aip,
 amsmath,amssymb,
preprint,%
]{revtex4-1}

\usepackage{graphicx}
\usepackage{dcolumn}
\usepackage{bm}
\usepackage{multirow} 
\DeclareUnicodeCharacter{2212}{-}
\begin{document}


\title{Evaluation of Structural Properties and Defect Energetics in Al$_x$Ga$_{1-x}$N Alloys} 



\author{Farshid Reza}
\affiliation{Department of Nuclear Engineering, The Pennsylvania State University, University Park, PA 16802, USA}
\author{Beihan Chen}
\affiliation{Department of Nuclear Engineering, The Pennsylvania State University, University Park, PA 16802, USA}
\author{Miaomiao Jin}
\email{mmjin@psu.edu (Corresponding author)}
\affiliation{Department of Nuclear Engineering, The Pennsylvania State University, University Park, PA 16802, USA}



\begin{abstract}
Al$_x$Ga$_{1-x}$N alloys are essential for high-performance optoelectronic and power devices, yet the role of composition on defect energetics remains underexplored, largely due to the limitations of first-principles methods in modeling disordered alloys. To address this, we employ a machine learning interatomic potential (MLIP) to investigate the structural and defect-related physical properties in Al$_x$Ga$_{1-x}$N. The MLIP is first validated by reproducing the equation of state, lattice constants, and elastic constants of the binary endpoints, GaN and AlN, as well as known defect formation and migration energies from density functional theory and empirical potentials. We then apply the MLIP to evaluate elastic constants of AlGaN alloys, which reveals a non-linear relation with alloying effect. Our results reveal that nitrogen Frenkel pair formation energies and the migration barriers for nitrogen point defects are highly sensitive to the local chemical environment and migration path. In contrast, Ga and Al vacancy migration energies remain relatively insensitive to alloy composition, whereas their interstitial migration energies exhibit stronger compositional dependence. These results provide quantitative insight into how alloying influences defect energetics in AlGaN, informing defect engineering strategies for improved material performance.
\end{abstract}
 

\maketitle 


\section{Introduction}
\label{sec:sample1}

Al$_x$Ga$_{1-x}$N alloys are widely used in high-frequency and high-power electronic devices, including high-electron mobility transistors (HEMTs), light-emitting diodes (LEDs), and RF amplifiers, due to their wide bandgap and high electron mobility \cite{dridi2002pressure, mishra2008gan, fletcher2017survey, piprek2012ultra, cciccek2024review}. Their large bandgap and polarization effects also make them well suited for extreme environments, such as space-based electronics, where radiation resistance is critical \cite{pearton2015ionizing}. However, various radiation exposures can still introduce defects, including vacancies, interstitials, and extended defect structures, that degrade electronic properties and device reliability \cite{liu1998ion, kucheyev2000ion, islam2019heavy, sequeira2021unravelling, kuboyama2011single}. Understanding how such defects form and migrate at the atomic scale is critical for predicting material performance under operational stresses.

Density functional theory (DFT) is widely used for calculating structural and defect properties with reasonable accuracy, such as formation energies and migration barriers \cite{lei2023comparative, kyrtsos2016migration}. However, DFT is computationally expensive and limited to small systems, which restricts its ability to model disordered alloys. Classical molecular dynamics (MD) simulations, enabled by empirical interatomic potentials such as the Tersoff and Stillinger–Weber (SW) models, can scale to larger sizes \cite{bockmann2002nanoseconds, nord2003molecular}, but often fail to deliver high-level accuracy, especially for defect formation and migration energetics \cite{gao2004intrinsic, lei2023comparative, kyrtsos2016migration}. Moreover, these empirical force fields are typically fitted to specific material compositions and lack generalizability to alloy systems with compositional disorder.

Although there has been significant work on GaN and AlN using both DFT and empirical MD, atomistic modeling of Al$_x$Ga$_{1-x}$N alloys has received much less attention. Prior MD study on the alloy investigated thermal conductance at different AlGaN interfaces \cite{huang2023first, luo2025heat}, interstitial and substitutional flourine ion movement in pure GaN and 25\% AlGaN \cite{yuan2008molecular}, and Al effect on defect production during irradiation \cite{jin2025examining}. Previous DFT studies on AlGaN include: vacancy diffusion in 30\% AlGaN under the influence of strain and electric field \cite{warnick2011room}, and vacancy state near valence and conduction band in Al$_6$Ga$_{24}$N$_{30}$ structure \cite{li2023electronic}. Although empirical force fields such as Tersoff and Stillinger–Weber (SW) have been developed for AlGaN systems \cite{zhou2013relationship, sun2021misfit, zhou2013molecular, karaaslan2020assessment}, they often lack the accuracy to capture the fine details of defect energetics. Consequently, the impact of alloy composition and local atomic fluctuations on defect formation and migration remains poorly understood. This knowledge gap is critical for predicting how AlGaN responds to radiation damage and thermal activation in real-world applications.

Recent advances in machine learning interatomic potentials (MLIPs) offer a promising pathway to overcome the limitations of both DFT and empirical force fields by combining near-DFT accuracy with the computational efficiency of classical MD. Among them, neural network potentials (NNPs) have demonstrated strong predictive power across a wide range of material systems, such as carbon \cite{wang2022deep}, Ag$_2$S \cite{balyakin2022deep}, GaN \cite{wu2024deep}, Ga$_2$O$_3$ \cite{li2020deep}, AlN \cite{li2024deep}, and AlGaN alloys \cite{huang2023first}. Despite these advancements, the capability of these models to resolve local chemical effects and capture composition-dependent defect behavior across the full composition range of Al$_x$Ga$_{1-x}$N alloys has yet to be utilized. 

In this work, we address this gap by using the MLIP to investigate the structural and defect-related physical properties of Al$_x$Ga$_{1-x}$N alloys. The capability of the potential is first benchmarked against DFT and literature data for the binary endpoints GaN and AlN, reproducing the equation of state, elastic constants, and point defect formation and migration energies. We then apply the model to alloy systems to examine how those properties vary with composition and local environment. These results offer new insights into the defect physics of AlGaN alloys.

\section{Computational Details}

All simulations in this study were conducted using the MLIP for AlGaN developed by Huang et al. \cite{huang2023first}. This NNP was trained on DFT data across a wide range of configurations and compositions. The potential was implemented in the LAMMPS molecular dynamics package \cite{thompson2022lammps} via integration with the DeepMD-kit framework \cite{wang2018deepmd}, enabling the use of ML-based force evaluations within classical MD simulations. A 2880-atom supercell based on the wurtzite crystal structure was constructed for GaN, AlN, and Al$_x$Ga$_{1-x}$N alloys with $x = 0.25$, $0.50$, and $0.75$. Periodic boundary conditions were applied in all directions. For static calculations, the supercells were modified based on the specific property being evaluated. For dynamics simulations, the temperature was controlled using a Nosé–Hoover thermostat at 300 K, unless otherwise noted, and the simulation timestep was set to 1 fs. 

To validate the ML potential, the equation of state (EOS) was computed for pure GaN and AlN. This was done by systematically varying the supercell volume through uniform expansion and contraction of lattice parameters. The energy–volume data were then fitted to the Birch-Murnaghan model to extract the equilibrium lattice parameters and minimum energy, which were compared with literature values to confirm consistency.

To assess whether the Al and Ga atoms in the Al$_x$Ga$_{1-x}$N alloys adopt any preferential ordering, Monte Carlo molecular dynamics (MCMD) simulations were performed for each alloy system. Simulations were carried out at 300 K for 100,000 MC steps. During these simulations, atomic swaps between Al and Ga atoms were allowed, and the total potential energy was monitored to evaluate whether the system relaxes into a chemically ordered or random configuration. Radial distribution functions (RDFs) were computed to detect any emergence of short-range order.

Furthermore, elastic constants were calculated at ground state for different AlGaN alloys. For the wurtzite structures: $C_{11}$, $C_{12}$, $C_{13}$, $C_{33}$, and $C_{44}$ are independent constants. These constants were determined by deformation of the simulation box along proper directions and by determining the change in the stress tensor \cite{elastic_constants}. The bulk modulus can then be obtained using the following equation:
\begin{equation}\label{eq: Bulk modulus}
    B=\frac{Y}{2(1+v)}
\end{equation}
Here, $Y$ is Young modulus, by $Y=\frac{(C_{11}-C_{12})(C_{11}+2C_{12})}{C_{11}+C_{12}}$ and $v$ is the Poisson ratio given by $v=\frac{C_{12}}{C_{11}+C_{12}}$ \cite{krieger1995elastic}.

To investigate defect behavior in Al$_x$Ga$_{1-x}$N, we introduced point defects into the relaxed 2880-atom perfect supercells. These include vacancies (V$_\text{Al}$, V$_\text{Ga}$, V$_\text{N}$), interstitials (Al$_i$ and Ga$_i$ in octahedral configuration, and N$_i$ in split configuration), and Frenkel pairs (a combination of a vacancy and an interstitial of the same species).  In this study, defect formation energy calculations were performed exclusively for Frenkel pairs in alloy systems, while in the binary compounds (GaN and AlN), we also examined Schottky defects, modeled by simultaneously removing one cation and one anion from the supercell for comparison with literature. All defective structures were energy-minimized at 0 K to obtain their relaxed configurations. The formation energy of a defect complex (Frenkel pair or Schottky defect), $E_f$, was calculated as the energetic cost of creating the defect from a perfect crystal, using the following expression:
\begin{equation}\label{eq: defect formation energy}
E_f = E^{\text{def}} - \frac{N_d}{N} E^{\text{perf}}
\end{equation}
where $E^{\text{def}}$ is the total energy of the defective supercell, $E^{\text{perf}}$ is the total energy of the defect-free supercell, $N_d$ is the number of atoms in the defective supercell, and $N$ is the number of atoms in the perfect supercell.  

Defect migration barriers were determined using the climbing-image nudged elastic band (CI-NEB) method \cite{henkelman2000improved, henkelman2000climbing, nakano2008space,maras2016global}. The initial and final configurations of the defect were constructed by translating the defect species along plausible migration paths inferred from literature studies on the binary nitrides (e.g., interstitialcy mechanism for interstitials \cite{kyrtsos2016migration, limpijumnong2004diffusivity, zhu2023computational} and nearest-neighbor hops for vacancies). Intermediate images were linearly interpolated and then relaxed with the NEB algorithm to trace the minimum energy path. The migration energy $E_m$ is defined as the energy difference between the highest-energy saddle point along the path and the initial configuration. All NEB calculations were also performed using the 2880-atom supercell to minimize finite-size effects and better capture local structural relaxation.

\section{Results and Discussion}

\subsection{Lattice Constant \& Elastic Properties}

\begin{figure}[!h]
    \centering
    \includegraphics[width=1.0\textwidth]{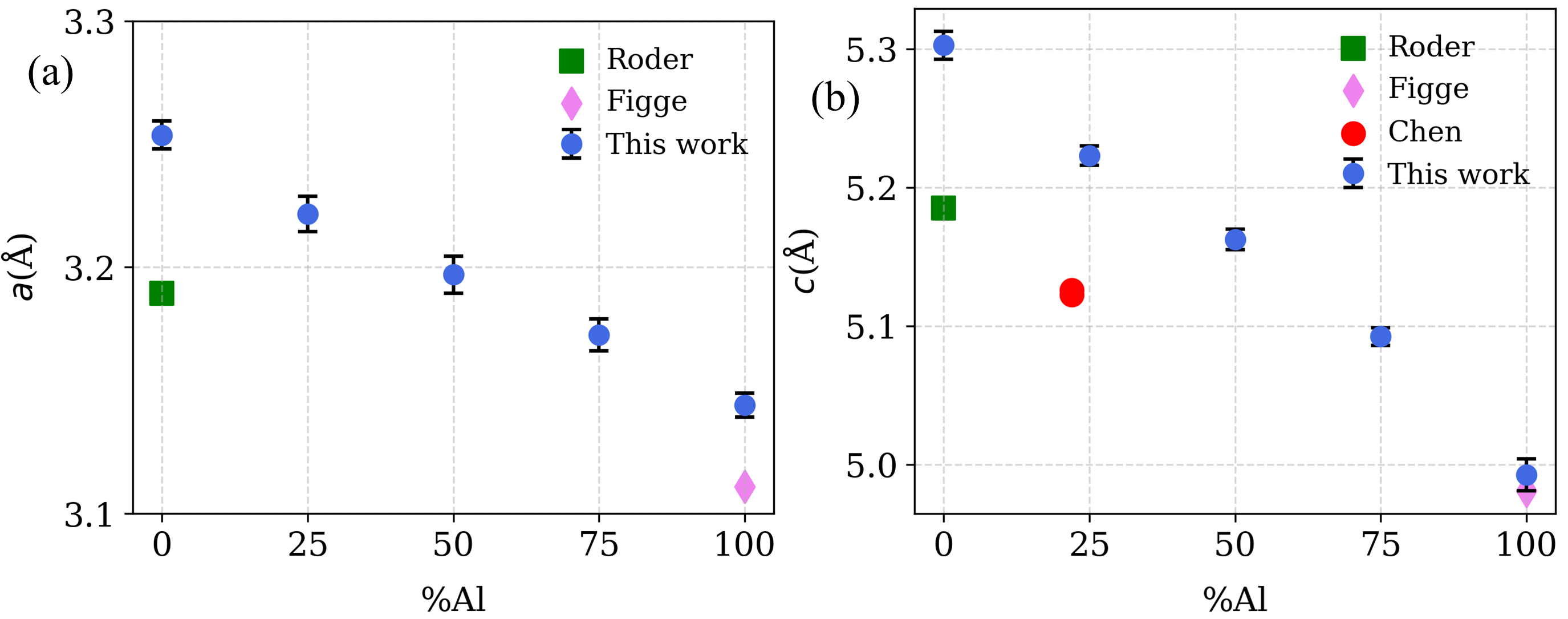}
    \caption {ML calculated lattice constant at 300 K for different Al$_x$Ga$_{1-x}$N alloys and comparison to previous experimental data by Roder et al. \cite{roder2005temperature}, Figge et al. \cite{figge2009temperature}, and Chen et al. \cite{chen2006high} along the $a$-direction (a) and $c$-direction (b).} 
    \label{fig:GaN_AlN_Lattice_constant}
\end{figure}

The EOS for the pure nitride systems was obtained using the MLIP to establish a reference for alloy analysis. MCMD simulations were then carried out for AlGaN alloys to examine potential short-range ordering. No evidence of short-range order was observed across the studied compositions. The MLIP accurately reproduces EOS trends consistent with prior literature, confirming its reliability for describing these systems. Additional details of the EOS and MCMD procedures are provided in the Supplementary Material (SM). Building on this validation, we extended the analysis to Al$_x$Ga$_{1-x}$N alloys for the equilibrium lattice constants at 300 K. As shown in Fig.~\ref{fig:GaN_AlN_Lattice_constant}, both the $a$- and $c$-lattice constants decrease monotonically with increasing Al content. This trend follows the Vegard’s law \cite{morales2013evaluation,ambacher2021wurtzite}. The predicted compositional trend aligns well with available experimental measurements. However, the ML-predicted lattice constants are consistently overestimated relative to the experimental data for GaN and AlN \cite{roder2005temperature, figge2009temperature}, as well as for Al$_{0.22}$Ga$_{0.78}$N reported by Chen et al. \cite{chen2006high}. This overestimation is a known artifact of the Perdew–Burke–Ernzerhof (PBE) exchange-correlation functional \cite{perdew1998perdew} used to generate the DFT training data for the ML model. PBE typically predicts slightly larger lattice constants than hybrid functionals or experiments \cite{gonzalez2014comparative}. Despite this systematic offset, the MLIP captures the relative variation of lattice constants across alloy compositions, which is critical for studying local strain responses due to defects.

Next, we examined the elastic properties to understand how alloying influences the mechanical response. The computed values of elastic constants ($C_{11}$, $C_{12}$, $C_{13}$, $C_{33}$, and $C_{44}$, and the derived $B$) are summarized in Table~\ref{tab: elastic constants for AlGaN alloys}. For the binary compounds GaN and AlN, the calculated elastic constants agree well with values reported in the literature \cite{shimada1998first, kim1996elastic, usman2011structural}. At low Al concentrations (e.g., Al$_{0.25}$Ga$_{0.75}$N), the introduction of smaller and lighter Al atoms into the GaN matrix leads to an obvious reduction in $C_{11}$, $C_{12}$, $C_{13}$, $C_{33}$, and $B$. This softening effect reflects the local lattice distortion and weakening of the bonding network as Al substitutes Ga, disrupting the native Ga–N bond environment. However, this relationship is not linear with composition. As the Al fraction increases further, $C_{12}$ and $C_{13}$ plateau, while $C_{11}$ begins to increase, meaning enhanced resistance to uniaxial deformation in Al-rich alloys along basal plane. $C_{33}$, which reflects the stiffness along the $c$-axis, is relatively high in both binary endpoints, but decreases noticeably in the intermediate alloys; this reduction suggests that alloying disrupts the strong directional bonding along the $c$-axis, possibly due to local lattice distortions. Regarding $C_{44}$, the increase across the entire composition range suggests that shear resistance improves with increasing Al content, potentially due to the stronger Al-N bond strength (Table~S1). Lastly, the bulk modulus exhibits a non-monotonic trend: it initially decreases with Al addition but partially recovers at high Al content.

\begin{table}[h!]
\centering
\caption{Elastic constants and bulk modulus of Al$_x$Ga$_{1-x}$N alloys from ML predictions, experiments, and DFT calculations (GPa).}
\label{tab: elastic constants for AlGaN alloys}
\begin{tabular}{|c|c|c|c|c|c|c|c|}
\hline
\textbf{Material} & \textbf{Source} & $C_{11}$ & $C_{12}$ & $C_{13}$ & $C_{33}$ & $C_{44}$ & $B$ \\
\hline
\multirow{3}{*}{GaN} 
 & This work & 374 & 183 & 148 & 378 & 85 & 247 \\
 & Experiment \cite{kim1996elastic} & 391 & 143 & 108 & 399 & 103 & 188–245 \\
 & DFT \cite{usman2011structural} & 329 & 109 & 80 & 357 & 91 & 176 \\
\hline
Al$_{0.25}$Ga$_{0.75}$N & This work & 338 & 128 & 97 & 340 & 93 & 198 \\
\hline
Al$_{0.5}$Ga$_{0.5}$N & This work & 355 & 126 & 97 & 344 & 101 & 204 \\
\hline
Al$_{0.75}$Ga$_{0.25}$N & This work & 366 & 128 & 97 & 331 & 108 & 207 \\
\hline
\multirow{3}{*}{AlN} 
 & This work & 377 & 132 & 98 & 368 & 116 & 213 \\
 & Experiment \cite{kim1996elastic} & 345 & 125 & 120 & 395 & 118 & 185–212 \\
 & DFT \cite{shimada1998first} & 398 & 142 & 112 & 383 & 127 & 195 \\
\hline
\end{tabular}
\end{table}

\subsection{Frenkel Pair Formation Energy}
\begin{figure}[!h]
    \centering
    \includegraphics[width=1.0\textwidth,scale=1.0]{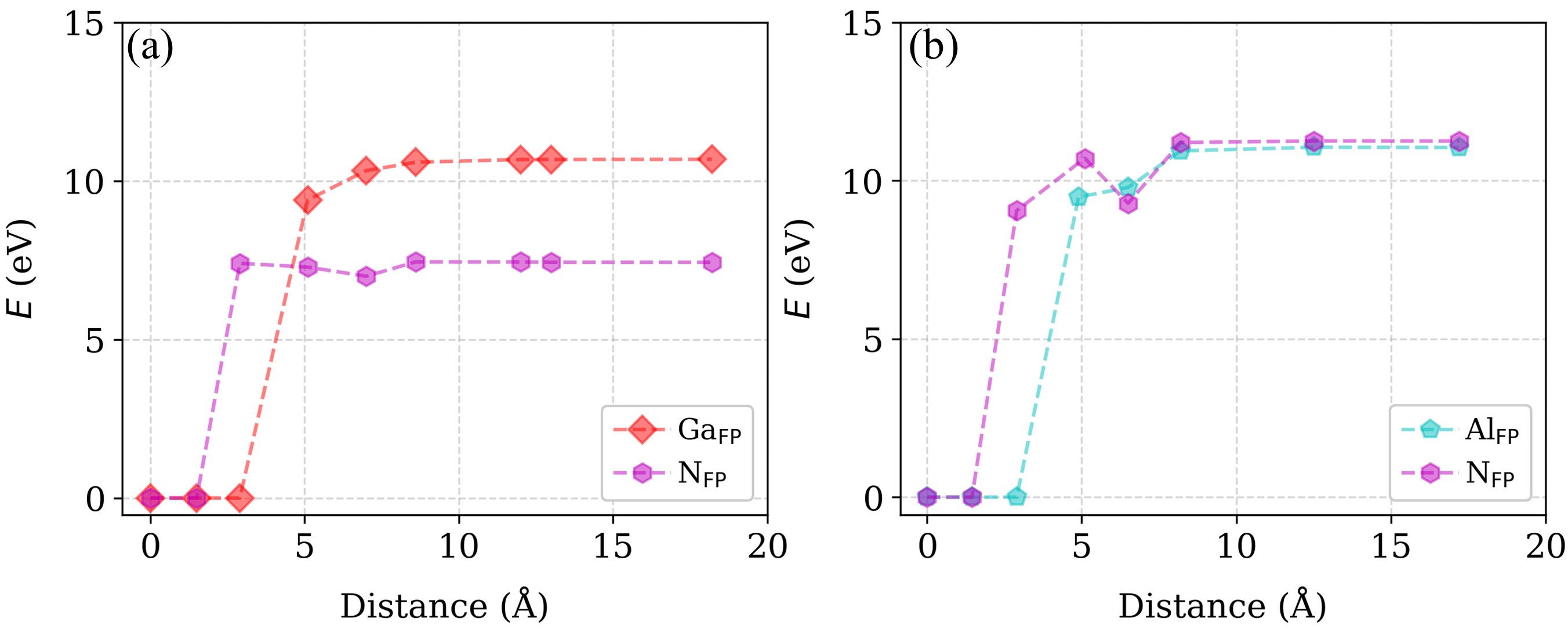}
    \caption {Frenkel pair interaction energy as a function of Frenkel pair distance in GaN (a) and AlN (b).} 
    \label{fig:Frenkel Pair vs Distance}
\end{figure}

Frenkel pair formation energy plays a critical role in understanding its response to radiation. We begin by benchmarking the formation energies in pure GaN and AlN. For Frenkel pairs, we first carried out formation energy calculations as a function of vacancy–interstitial separation. This was to identify a separation at which the interstitial does not spontaneously recombine with the vacancy during relaxation. For each case, the interstitial was displaced incrementally from its corresponding vacancy site, and the total energy was minimized to obtain the system energy with respect to the energy of the perfect supercell. As shown in Fig.~\ref{fig:Frenkel Pair vs Distance}, stable Frenkel pair formation for Ga and Al occurs when the vacancy–interstitial distance exceeds 5 {\AA}. Below this, the defect pair annihilates. In contrast, N Frenkel pair becomes stable at short separation ($\sim$3 {\AA}). It can also be seen that in GaN, the formation energy of N Frenkel pairs is significantly lower than that of Ga Frenkel pairs, whereas in AlN, the formation energies for N and Al Frenkel pairs are comparable. The quantitative formation energies for well-separated defects are summarized in Table~\ref{table:defect_data}, and show strong consistency with previously reported DFT calculations for GaN and empirical potential results for AlN (no ab initio data for Frenkel or Schottky defects in AlN are currently available in the literature).  

\begin{table}[h]
    \centering
    \caption{Defect formation energies (in eV) in GaN and AlN.}
    \label{table:defect_data}
    \begin{tabular}{|c|c|c|c|c|}
        \hline
        \textrm{Material} & \textrm{Defect Type} & \textrm{This work} & \textrm{References} \\
        \hline
        \multirow{3}{*}{GaN} 
            & Ga$_\text{FP}$ & 10.68 & 10.07 \cite{lei2023comparative}  \\
            & N$_\text{FP}$ & 7.43 & 7.32 \cite{lei2023comparative} \\
            & V$_\text{GaN}$ Schottky & 6.42 & 6.66 \cite{lei2023comparative}\\
        \hline
        \multirow{3}{*}{AlN} 
            & Al$_\text{FP}$ & 11.05 & 10.47 \cite{zhu2023computational}  \\
            & N$_\text{FP}$ & 11.25 & 10.52 \cite{zhu2023computational}  \\
            & V$_\text{AlN}$ Schottky & 6.06 & 8.16 \cite{zhu2023computational}  \\
        \hline
    \end{tabular}
\end{table}

In addition, we also computed the formation energy of Schottky defects, which are created by removing a pair of nearest-neighbor cation and anion atoms along the basal plane, consistent with prior studies \cite{lei2023comparative,zhu2023computational}. The formation energy was calculated using Eq. (\ref{eq: defect formation energy}), with results presented in Table~\ref{table:defect_data}.  For GaN, our results closely match the DFT data reported by Lei et al. \cite{lei2023comparative}. For AlN, in the absence of ab initio data, we compare our results to those obtained using the empirical force field developed by Zhu et al. \cite{zhu2023computational}, and note a difference of 2.1 eV. These results overall demonstrate that the AlGaN ML potential can reliably capture defect formation energies in the binary compounds.


We next investigated how alloying affects Frenkel pair formation in Al$_x$Ga$_{1-x}$N. Unlike pure binaries, alloys present a variety of local atomic configurations due to random distribution of Ga and Al atoms on the cation sublattice. To capture this configurational variability, 100 random atomic configurations were generated for each alloy composition. In each configuration, Frenkel pairs for Ga, Al, and N were individually introduced to be well separated vacancy-interstitial pair, and then the system was fully relaxed to obtain defect formation energies.

\begin{figure}[!h]
    \centering
    \includegraphics[width=0.6\textwidth,scale=1.0]{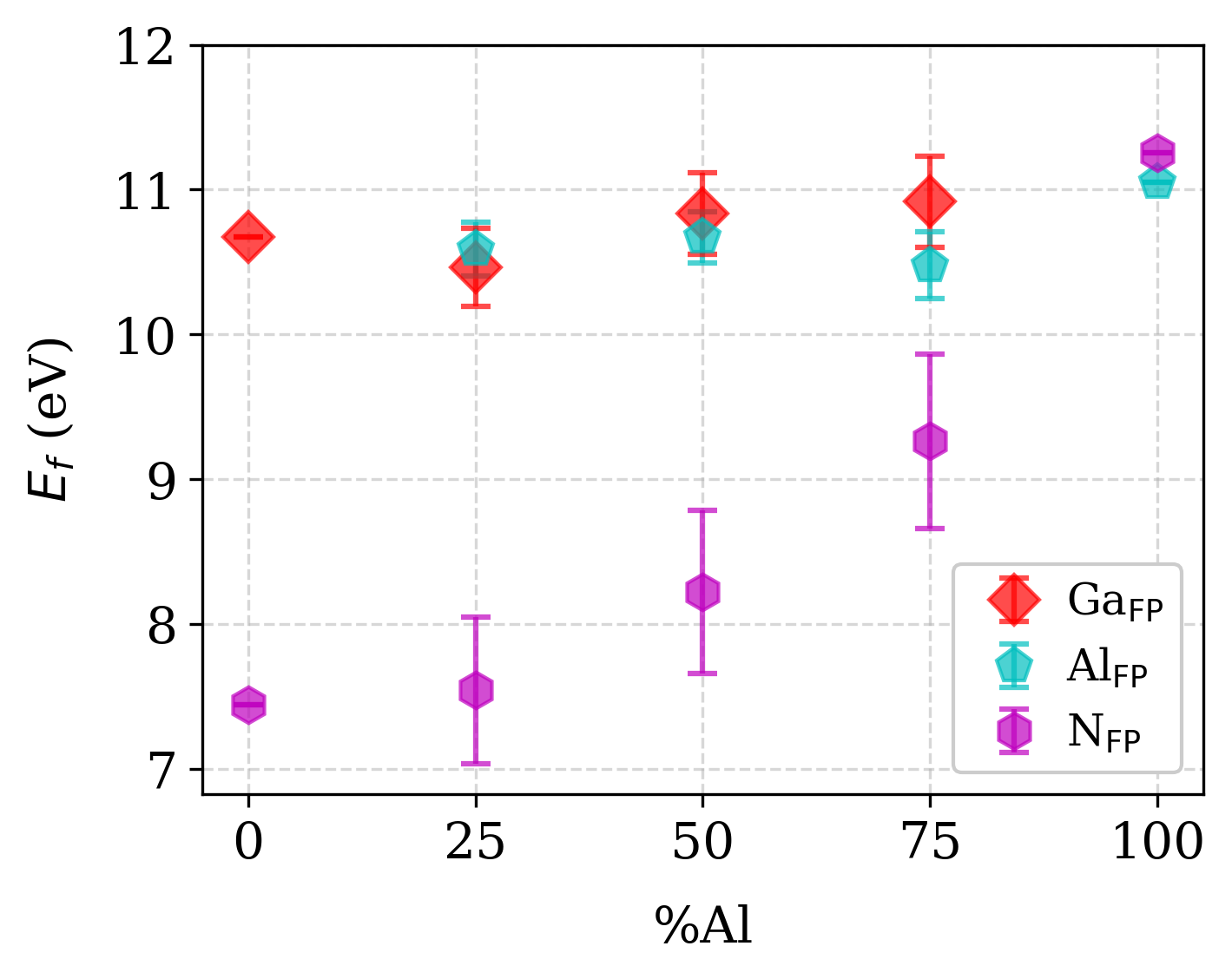}
    \caption {Frenkel pair formation energy as a function of Al percentage.} 
    \label{fig:concentration frenkel}
\end{figure}

\begin{figure}[!h]
    \centering
    \includegraphics[width=0.6\textwidth,scale=1.0]{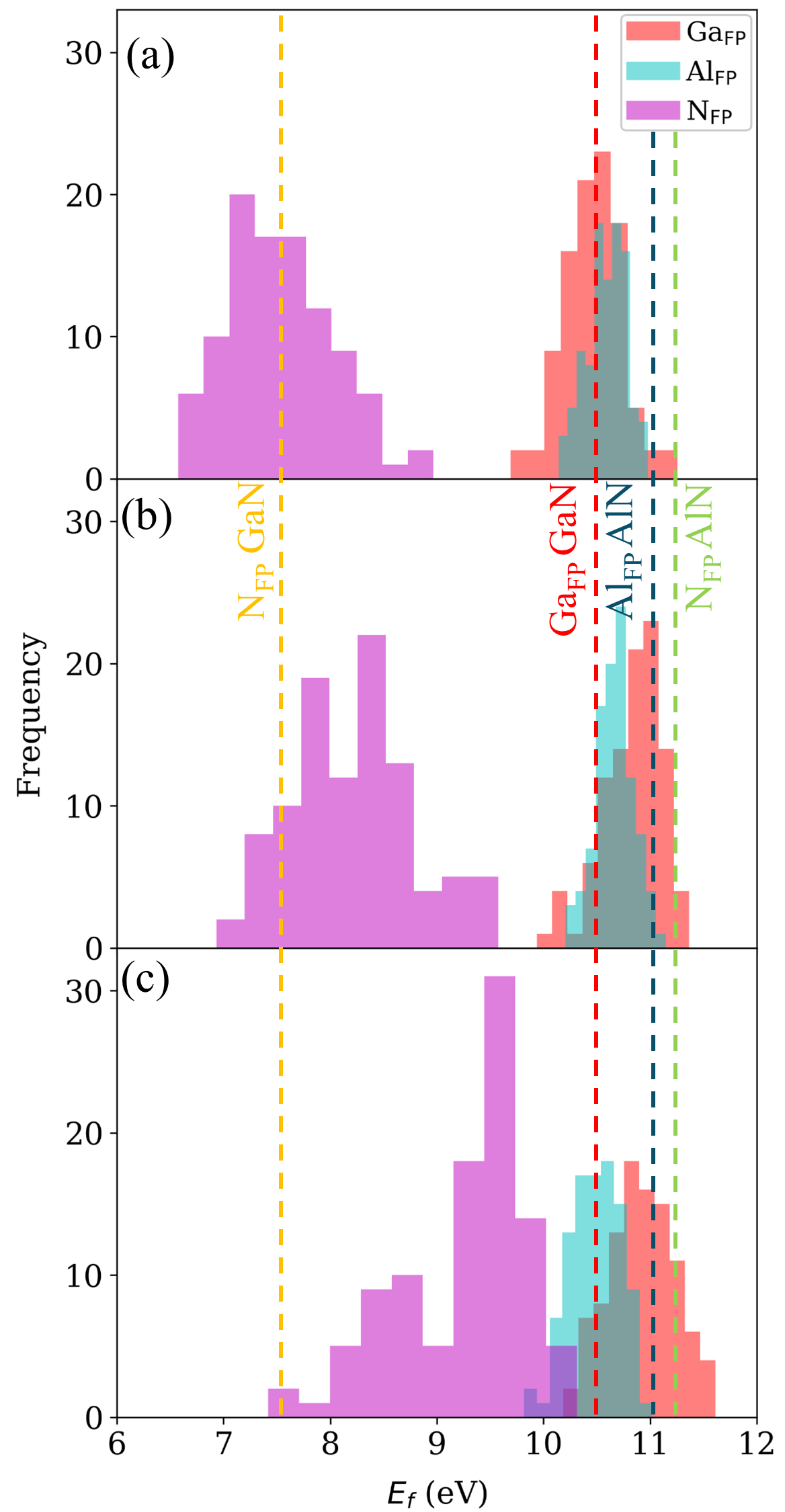}
    \caption {Histogram of the Frenkel pair formation energies in 25\% AlGaN (a), 50\% AlGaN (b), and 75\% AlGaN (c). As a reference, dashed lines indicate the corresponding Frenkel pair energy in the pure nitrides.} 
    \label{fig:frenkel pair energy distribution in alloys}
\end{figure}

Fig.~\ref{fig:concentration frenkel} summarizes the results for all compositions. For Ga and Al Frenkel pairs, the average formation energy exhibits only a weak dependence on alloy composition, and the spread in formation energies remains relatively small across all alloys. It can be attributed to the similar atomic sizes and bonding environments of Ga and Al cations and the relatively large interstitial sites they occupy, which leads to similar vacancy and interstitial formation energies. In contrast, the behavior of N Frenkel pairs is strongly influenced by alloy composition. Both the average formation energy and the standard deviation increase significantly with rising Al content. This is primarily attributed to the stronger Al–N bonds, which raise the energy cost of defect formation. The large standard deviation stems from increased variation in bonding strength and local strain fields.

To gain a deeper understanding of the configurational effects of alloying on Frenkel pair energetics, Fig.~\ref{fig:frenkel pair energy distribution in alloys} shows the formation energy distribution of Frenkel pairs across the different alloy compositions, with the formation energies for the corresponding pure GaN and AlN systems indicated by dashed vertical lines.  For Ga and Al Frenkel pairs, the energy distributions remain narrow and show minimal compositional dependence, consistent with the earlier observation that cation Frenkel energetics are largely insensitive to alloying. The minimum formation energy of both cation Frenkel pairs is around 10 eV, which is lower than that in their pure nitrides. 

For N Frenkel pair, the energy distributions exhibit pronounced broadening and a shift toward higher energies with increasing Al content. At 75\% Al, the N Frenkel pair energy distribution exhibits a bimodal character, suggesting the emergence of two classes of local defect environments. This behavior likely reflects the increasing likelihood of N atoms being fully coordinated by Al neighbors due to high Al content, which results in high formation energies of N vacancies. Another observation is the low-energy N Frenkel pair configurations in the alloy. This is most evident at 25\% Al composition, where the energy histogram develops a `fat' low-energy tail. Representative configurations of such low energy instances are shown in the SM S3: the lowest-energy N Frenkel pair (6.57 eV) is found in a local environment enriched with Al neighbors around the N interstitial (SM Figure~S3a), while the highest-energy configuration (8.96 eV) occurs when the interstitial is primarily surrounded by Ga atoms (SM Figure~S3b), resulting in an energy difference of 2.39 eV. It reveals that local Al enrichment near the defect can impact N Frenkel pairs by providing shorter, stronger Al–N bonds. Interestingly, this stabilization effect decreases at higher Al concentrations (50\% and 75\%), where the N Frenkel energy distribution shifts upward overall and the low-energy tail starts to `thin'. This reflects a non-linear dependence of defect energetics on alloy composition: at low Al content, isolated Al atoms can locally soften the lattice around N defects reducing the energy penalty in forming defects, while at higher Al content, the dominance of strong Al–N bonding raises the overall formation energy. 

These results have implications in predicting defect behavior in AlGaN alloys. First, the presence of low-energy N Frenkel configurations in the alloy implies that under irradiation or during high-temperature processing, these configurations may form preferentially and dominate the early defect population, and this would influence the subsequent defect clustering and migration kinetics. Second, the compositional dependence suggests that careful control of local alloy composition such as through compositional grading could provide a pathway to tune defect tolerance in AlGaN-based devices.

\subsection{Migration Energy}

While defect formation energy determines the likelihood of defect generation, the migration energy of a defect governs its mobility to form complexes or clusters, which ultimately affects long-term material stability under thermal or irradiation conditions.  A number of studies have reported migration energies of native point defects in pure GaN and AlN systems using both DFT and empirical methods \cite{kyrtsos2016migration, limpijumnong2004diffusivity, zhu2023computational, hrytsak2021dft, li2024deep}. However, there is presently no study examining how defect migration energies evolve with alloy composition in Al$_x$Ga$_{1-x}$N. Given the pronounced sensitivity of N Frenkel pair formation energies to local chemical environments observed in this study, we expect that migration barriers may also exhibit significant composition and configuration dependence in the alloy. To address this gap, we investigated the migration energies of point defects (vacancies and interstitials) for different Al$_x$Ga$_{1-x}$N compositions.
\subsubsection{Vacancy Migration Energy}

We first evaluate vacancy migration energies in pure GaN and AlN. In the wurtzite structure, vacancy migration typically occurs along two paths: in-plane migration (perpendicular to the c-axis) and out-of-plane migration (parallel to the c-axis) \cite{warnick2011room}. Due to the anisotropic bonding in wurtzite nitrides, these two directions exhibit different migration barriers, as previously studied in DFT and experimental studies \cite{warnick2011room, tuomisto2007introduction}. Also, defect migration barriers are known to depend on defect charge states \cite{hrytsak2021dft}; however, since the current MLIP does not explicitly model charge effects, all results reported here correspond to neutral defects.

\begin{table}[h]
    \centering
    \caption{Comparison of in-plane and out-of-plane neutral vacancy migration energy values (in eV) in  GaN and AlN.}
   \label{tab:migration energy table}
    \begin{tabular}{|c|c|cc|cc|}
        \hline
        Material & Defect type & \multicolumn{2}{c|}{In-plane} & \multicolumn{2}{c|}{Out-of-plane} \\
        \cline{3-6}
                 &             & This work & References & This work & References \\
        \hline

        \multirow{2}{*}{GaN} 
            & V$_\text{Ga}$ & 1.88 & \begin{tabular}[c]{@{}l@{}}1.90 \cite{warnick2011room} \\  2.38 \cite{hrytsak2021dft} \\ 2.5 \cite{kyrtsos2016migration} \end{tabular}
                     & 2.49 & \begin{tabular}[c]{@{}l@{}}2.75 \cite{hrytsak2021dft} \\ 2.8 \cite{kyrtsos2016migration} \end{tabular} \\
        \cline{2-6}
            & V$_\text{N}$  & 2.48 & \begin{tabular}[c]{@{}l@{}}3.13 \cite{hrytsak2021dft} \\ 2.7 \cite{kyrtsos2016migration} \\ 2.0 \cite{warnick2011room} \end{tabular} 
                     & 3.27 & \begin{tabular}[c]{@{}l@{}}3.10 \cite{gao2019point} \\ 4.06 \cite{hrytsak2021dft} \\ 3.4 \cite{kyrtsos2016migration} \end{tabular} \\
        \hline

        \multirow{2}{*}{AlN} 
            & V$_\text{Al}$ & 2.23 & 2.37 \cite{li2024deep} & 2.76 & 2.97 \cite{li2024deep} \\
        \cline{2-6}
            & V$_\text{N}$  & 2.69 & 2.78 \cite{li2024deep} & 3.12 & \begin{tabular}[c]{@{}l@{}}3.5 \cite{li2024deep} \\ 3.37 \cite{gao2019point} \end{tabular} \\
        \hline
    \end{tabular}
\end{table}

Table~\ref{tab:migration energy table} summarizes our calculated vacancy migration energies in pure GaN and AlN, compared with available literature data on neutral vacancies. For GaN, the literature reports a range of migration energies, due to variations in computational approaches and approximations. For Ga vacancies, our ML-predicted in-plane migration energy (1.88 eV) is close to the DFT-reported range of 1.9–2.5 eV \cite{warnick2011room, kyrtsos2016migration, hrytsak2021dft}. The out-of-plane migration energy 2.49 eV likewise agree with DFT values (2.75-2.8 eV). For N vacancies, our in-plane migration energy (2.48 eV) is within the range of DFT-reported values (2.0–3.1 eV), with out-of-plane migration energies also showing close agreement. For AlN, the MLIP accurately reproduces DFT-calculated migration energies. The in-plane migration energy differs by only $\sim$0.1 eV from prior DFT values, while out-of-plane migration differs by up to $\sim$0.2 eV. Overall, these results confirm that the MLIP can reliably reproduce vacancy migration energies and their directional anisotropy in both pure GaN and AlN.

\begin{figure}[!h]
    \centering
    \includegraphics[width=1.0\textwidth,scale=1.0]{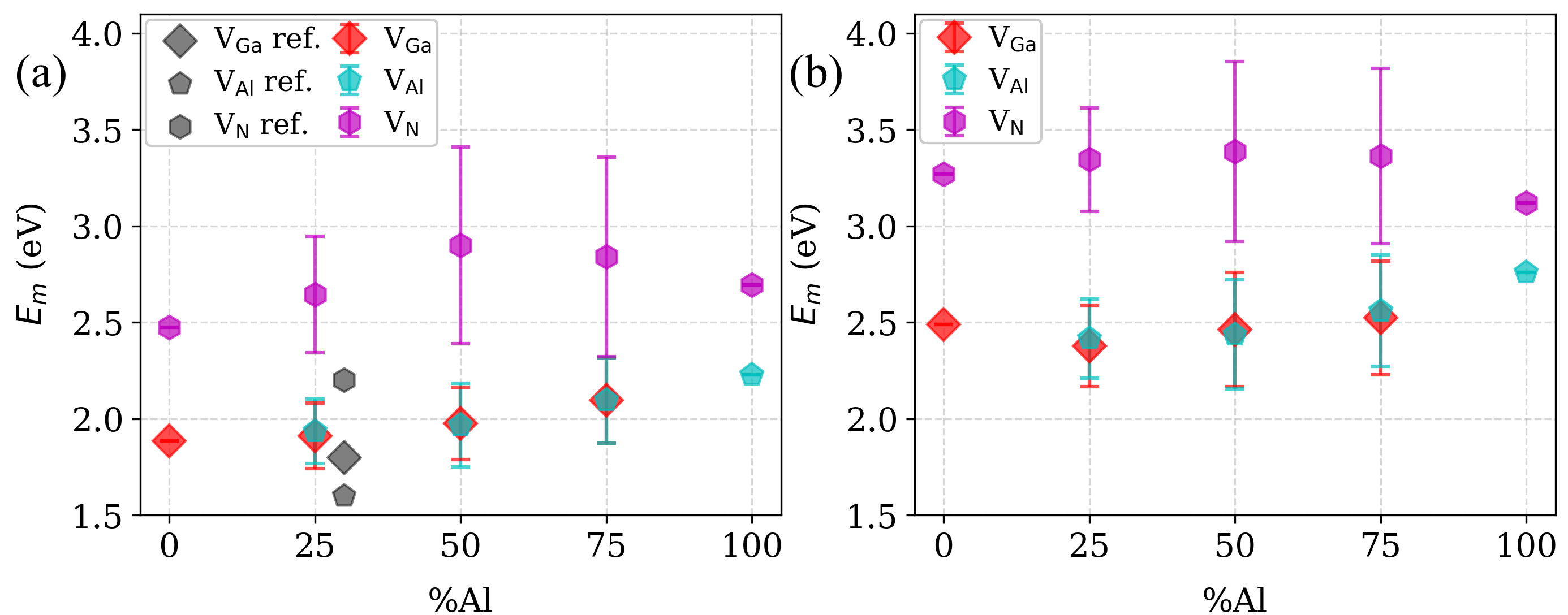}
    \caption {In-plane average vacancy migration energy with literature data of 30\% AlGaN \cite{warnick2011room} (a) and out-of-plane (b) average vacancy migration energy as function of Al\%.} 
    \label{fig:AlGaN vacancy migration comparison}
\end{figure}

\begin{figure}[!h]
    \centering
    \includegraphics[width=1.0\textwidth,scale=1.0]{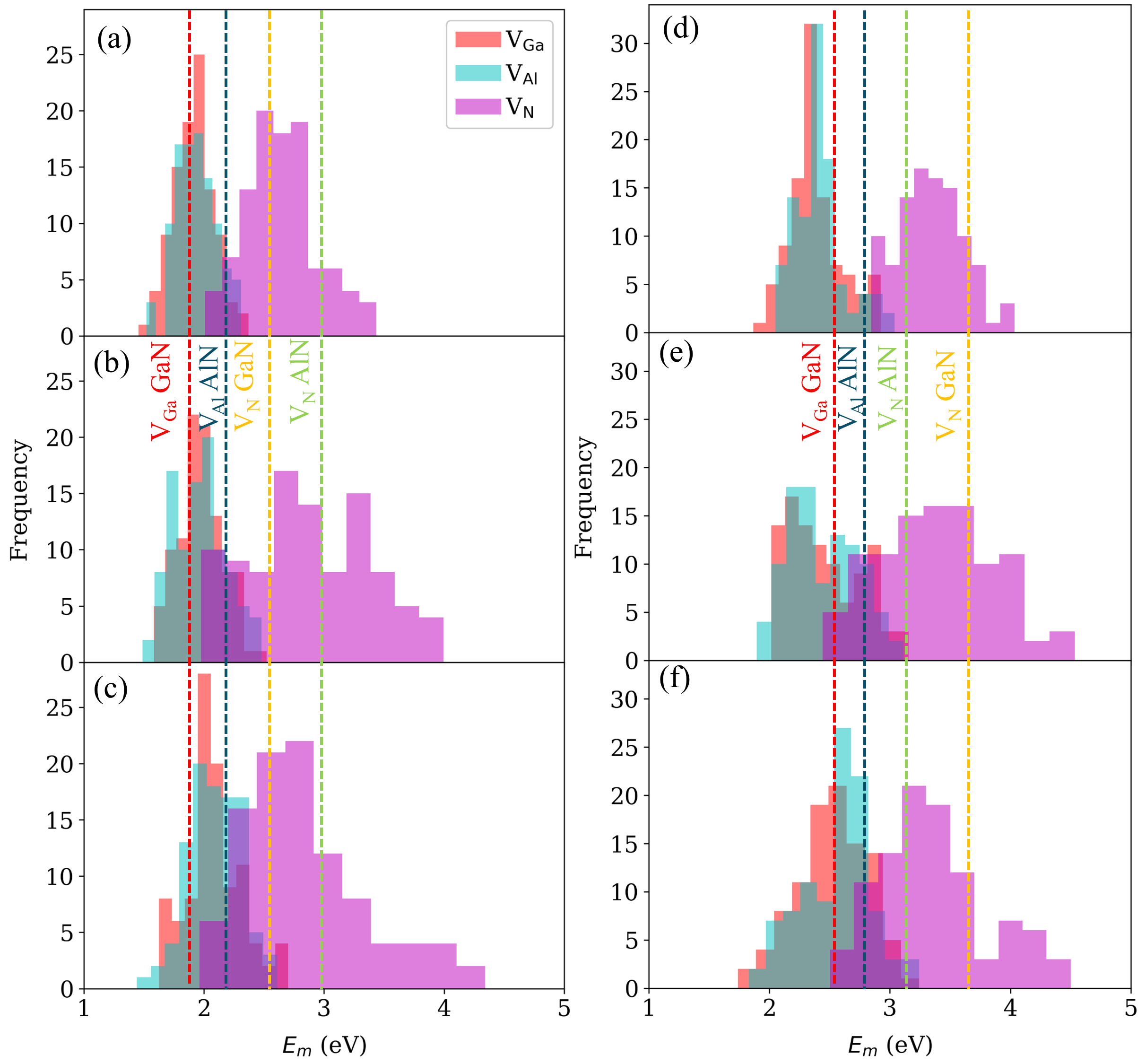}
    \caption {In-plane: 25\% AlGaN (a), 50\% AlGaN (b), 75\% AlGaN (c), and out-of-plane: 25\% AlGaN (d), 50\% AlGaN (e), 75\% AlGaN (f) vacancy migration energy statistics in AlGaN. As a reference, dashed lines indicate the corresponding vacancy migration energy in the pure nitrides.} 
    \label{fig:AlGaN migration statistics}
\end{figure}

Moving on to Al$_x$Ga$_{1-x}$N alloys, for each composition, we generated 100 random alloy configurations to capture the impact of local chemical variations on both in-plane and out-of-plane migration paths. The results are summarized in Fig.~\ref{fig:AlGaN vacancy migration comparison} for in-plane vacancy migration in Fig.~\ref{fig:AlGaN vacancy migration comparison}(a) and out-of-plane vacancy migration in Fig.~\ref{fig:AlGaN vacancy migration comparison}(b). There is a slight increase in the average vacancy migration energy for cation vacancies as Al content increases. This trend is consistent across both migration directions, although out-of-plane migration barriers remain higher than in-plane barriers at all compositions, consistent with the behavior in pure GaN and AlN (Table~\ref{tab:migration energy table}). For cation vacancies, the overall spread of vacancy migration energies remains moderate, with in-plane migration exhibiting a narrower range of variation compared to out-of-plane migration across all AlGaN alloy compositions (Fig.~\ref{fig:AlGaN migration statistics}). These trends indicate that cation vacancy migration remains relatively insensitive to local chemical fluctuations. As a comparison, previous DFT calculations by Warnick et al. \cite{warnick2011room} for Al$_{0.3}$Ga$_{0.7}$N reported in-plane migration barriers of 1.8 eV for Ga vacancies and 1.6 eV for Al vacancies in neutral states, which fall within the lower end of our computed in-plane migration energy spectrum for 25\% Al (Fig.~\ref{fig:AlGaN migration statistics}a). 

In contrast, average N vacancy migration energy peaks at 50\% Al, and the barriers show a much larger spread  (Fig.~\ref{fig:AlGaN migration statistics}), particularly in high Al content systems. It is worth noting that previous DFT calculations reported a N vacancy migration energy of 2.2 eV \cite{warnick2011room} for Al$_{0.3}$Ga$_{0.7}$N, which falls within the range of the current results (Fig.~\ref{fig:AlGaN migration statistics}a). These findings suggest that the local environments have large impact on N migration, and the sluggish diffusion is the most significant when local configurational disorder is greatest. At the same time, the presence of low-energy migration pathways within the alloy indicates that preferential diffusion channels may exist, enabling localized N vacancy transport in compositionally complex environments. At lower Al content (25\%), most migration paths still resemble Ga-rich environments. At 50\% Al, the local cation environments around N vacancies are highly mixed, with variation in Ga–N and Al–N bonding along the migration path, which leads to a higher average barrier. At 75\% Al, although the spread in migration barriers remains large, the average barrier slightly decreases as Al-rich environments become more dominant. This composition-dependent behavior underscores the sensitivity of N vacancy migration to the local Al/Ga arrangement along the migration path.

\subsubsection{Interstitial Migration Energy}

\begin{table}[h]
    \centering
    \caption{Comparison of interstitial migration energy values (in eV) in pure GaN and AlN (The charge state is indicated inside parentheses except for the neutral cases).}
   \label{tab:interstitial migration energy table}
    \begin{tabular}{|c|c|cc|cc|}
        \hline
        Material & Defect type & \multicolumn{2}{c|}{Interstitialcy} \\
        \cline{3-4}
                 &             & This work & References  \\
        \hline

        \multirow{2}{*}{GaN} 
            & Ga$_\text{i}$ & 0.85 & \begin{tabular}[c]{@{}l@{}}0.7(+3) \cite{kyrtsos2016migration} \\  No neutral state \\ 0.9(+3) \cite{limpijumnong2004diffusivity} \end{tabular}
                      \\
        \cline{2-4}
            & N$_\text{i}$  & 1.12 & \begin{tabular}[c]{@{}l@{}}2.4 \cite{kyrtsos2016migration} \\ 1.4 \cite{zhu2023computational} \end{tabular} 
                     \\
        \hline

        \multirow{2}{*}{AlN} 
            & Al$_\text{i}$ & 1.14 & 0.93(+3) \cite{zhu2023computational}  \\
        \cline{2-4}
            & N$_\text{i}$  & 1.46 & 1.32(-3) \cite{zhu2023computational}  \\
        \hline
    \end{tabular}
\end{table}

\begin{figure}[!h]
    \centering
    \includegraphics[width=0.6\textwidth,scale=1.0]{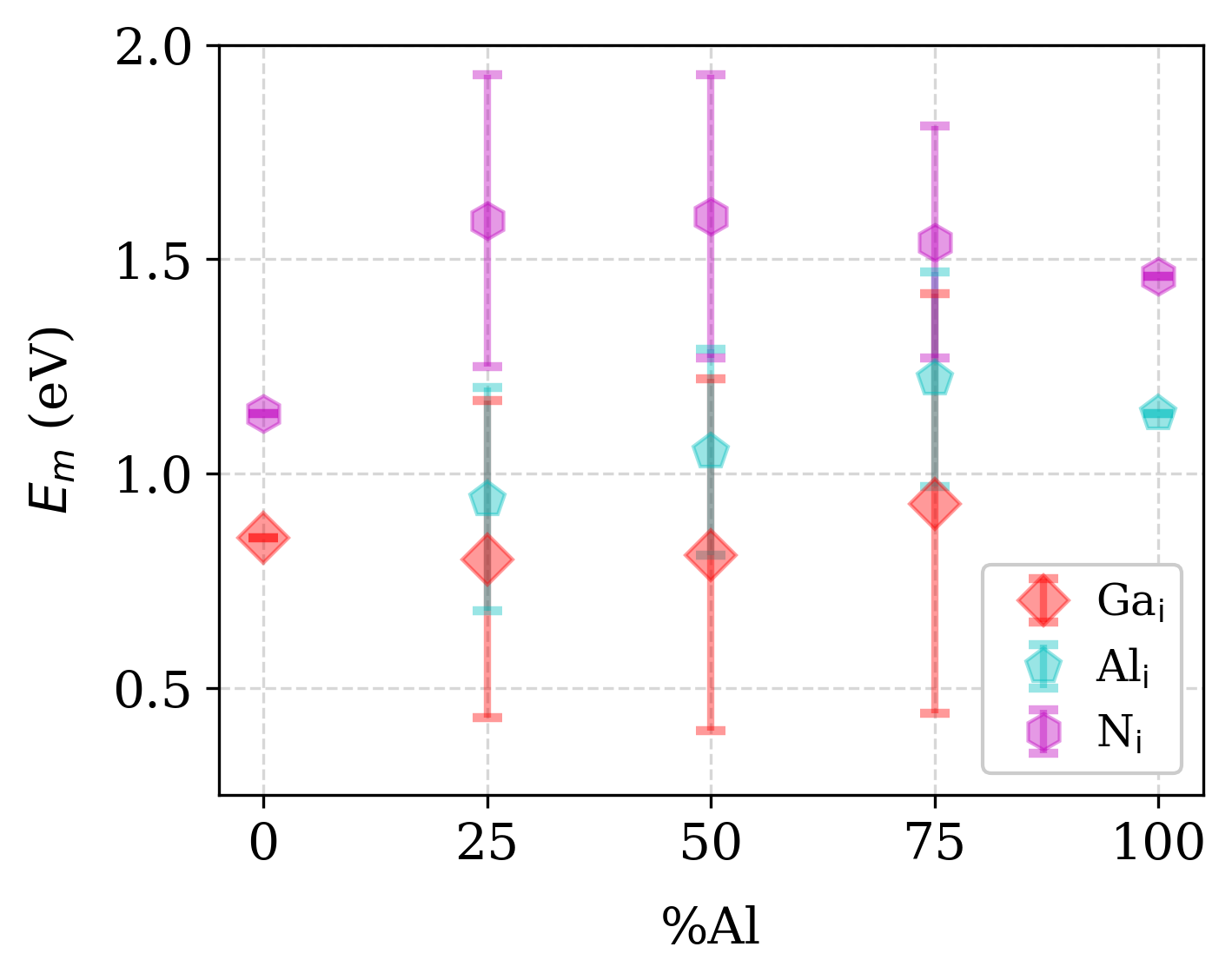}
    \caption {Interstitial migration energy barrier as function of Al\% via the interstitialcy mechanism.} 
    \label{fig:AlGaN interstitial migration comparison}
\end{figure}

\begin{figure}[!h]
    \centering
    \includegraphics[width=0.6\textwidth,scale=1.0]{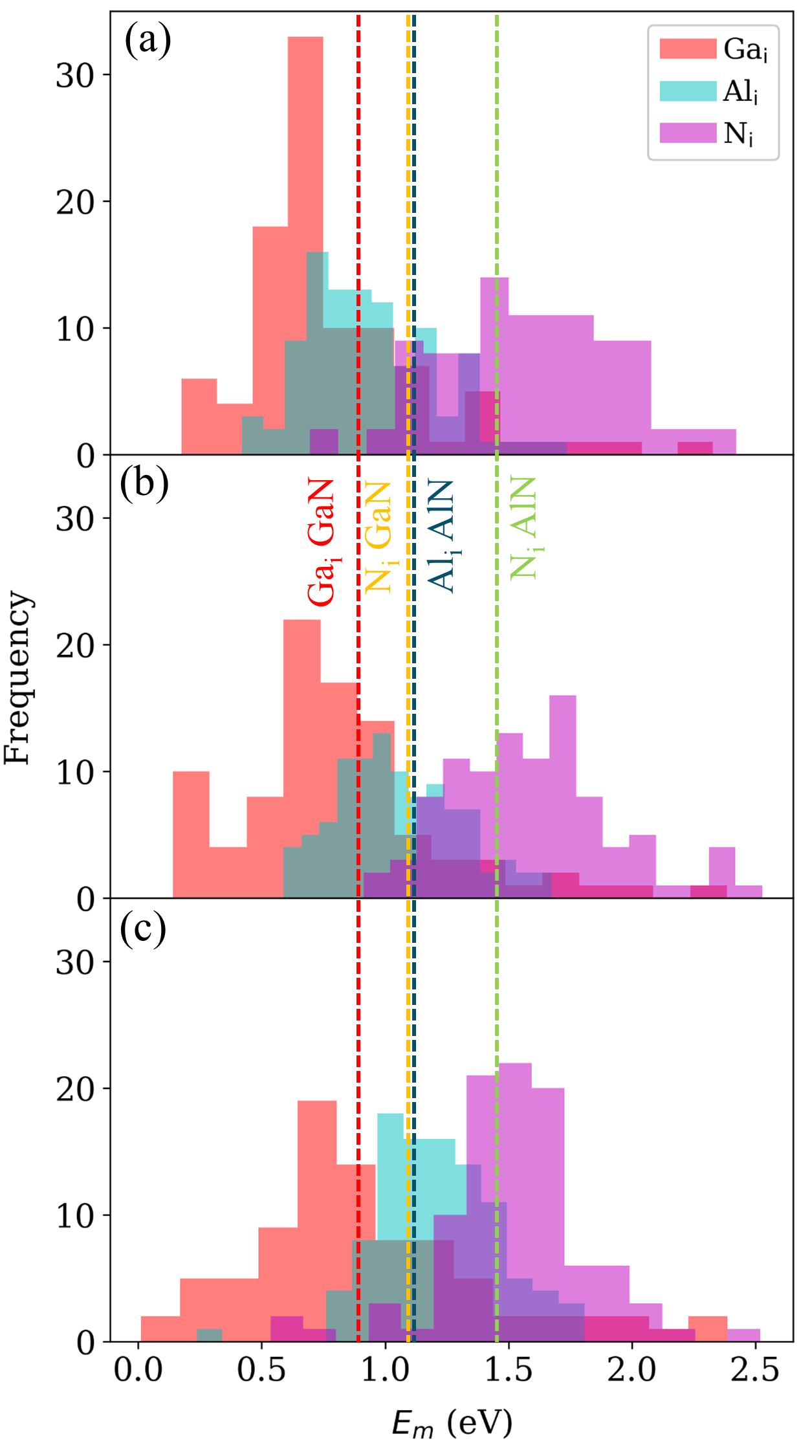}
    \caption {Distribution of interstitial migration energy via the interstitialcy mechanism in 25\% (a), 50\% (b), and 75\% Al (c). As a reference, dashed lines indicate the corresponding interstitial migration energy in the pure nitrides.} 
    \label{fig:AlGaN interstitial migration histogram}
\end{figure}

Regarding interstitial migration, existing work on pure GaN and AlN has identified two main migration mechanisms for interstitials: $c$-channel migration and the interstitialcy mechanism \cite{kyrtsos2016migration, hrytsak2021dft, limpijumnong2004diffusivity}. The literature indicates that the interstitialcy mechanism is more energy favorable (lower migration barriers than $c$-channel migration) \cite{kyrtsos2016migration}. Therefore, our study focused on the interstitialcy mechanism. In the interstitialcy mechanism, the interstitial atom displaces a host atom from its lattice site, occupying that site and pushing the host atom into an interstitial position.

We again first benchmarked the MLIP by calculating neutral-state interstitialcy migration barriers in pure GaN and AlN, comparing our results to the available literature data. The results are summarized in Table~\ref{tab:interstitial migration energy table}. For Ga interstitials in GaN, the calculated neutral-state migration barrier via the interstitialcy mechanism is 0.85 eV. Prior studies reported +3 charge state barriers (0.7 eV \cite{kyrtsos2016migration} and 0.9 eV \cite{limpijumnong2004diffusivity}). For N interstitials in GaN, the ML potential yields a barrier of 1.12 eV, which is close to the lower end of the reported range of neutral-state barriers (1.4–2.4 eV) from the literature \cite{kyrtsos2016migration, zhu2023computational}. In AlN, the ML-calculated neutral-state interstitial migration barriers show good agreement with prior empirical results reported for the +3 charge state of Al interstitials and the -3 charge state of N interstitials (Table~\ref{tab:interstitial migration energy table}). While the ML potential does not explicitly model charged defects, this comparison indicates that the migration pathways and relative energy scales captured by the ML model are physically realistic, even in the absence of explicit charge effects.

Next, we analyzed how migration barriers evolve in Al$_x$Ga$_{1-x}$N alloys. The average migration energy for Ga, Al, and N interstitials as a function of Al content is shown in Fig.~\ref{fig:AlGaN interstitial migration comparison}. For Ga interstitials, the average migration barrier exhibits only a modest change with alloy composition, while the overall spread of migration energies becomes significant in alloys (see Fig.~\ref{fig:AlGaN interstitial migration histogram} for the energy distribution). Hence, the migration of Ga interstitials is sensitive to local chemical environments. An interesting trend emerges at higher Al content: at 75\% Al, low-barrier Ga migration pathways appear. This tail of low-energy events implies that percolative Ga diffusion channels may develop in Al-rich alloys, potentially facilitating long-range Ga transport in these compositions. In contrast, Al interstitial migration barriers increase with Al content in alloys. In addition, the distribution of migration energies for Al interstitials becomes broader and shifts toward higher values as Al content increases (Fig.~\ref{fig:AlGaN interstitial migration histogram}), which stems from increasing stiffer Al–N bonding network on interstitial mobility. The behavior of N interstitials is particularly nuanced. The average migration energy for split N interstitials first increases by $\sim$0.4 eV when moving from pure GaN to 25\% Al. Between 25\% and 75\% Al, the average migration barrier remains relatively stable, but the histogram peak narrows (Fig.~\ref{fig:AlGaN interstitial migration histogram}), indicating that the probable migration pathways become more constrained as the alloy becomes Al-rich. Interestingly, at 75\% Al, the energy distribution also develops a tail of very low-barrier migration pathways, similar to the trend observed for Ga interstitials (Fig.~\ref{fig:AlGaN interstitial migration histogram}c). This suggests that in an Al-rich matrix, preferential low-energy diffusion channels for N interstitials may emerge.

\section{Conclusion}

In this study, we systematically applied MLIP to investigate the physical properties of Al$_x$Ga$_{1-x}$N alloys across a broad composition range, with a particular focus on their defect properties. This also represents the first effort to explore composition-dependent defect energetics in AlGaN using a high-fidelity ML potential. In those calculation workflows, the ML potential was first validated against experimental and first principles data. It well reproduces the equation of state, elastic constants, and defect formation and migration energies for the binary endpoints GaN and AlN. The major focus lies in extending these analyses to defect properties in AlGaN alloys, where conventional first-principles methods are computationally prohibitive due to the large configurational space. Our results reveal that both formation energy distributions and migration energy distributions in AlGaN alloys can be highly sensitive to local environments, especially for N defects. Notably, due to the alloying effect, the migration barriers of interstitials can be significantly reduced compared to those of pure nitrides, and the appearance of such low-barrier diffusion pathways in Al-rich alloys could lead to enhanced localized diffusion. Overall, this work provides atomistic insights of defect behavior in AlGaN alloys. These insights can inform the design of AlGaN-based optoelectronic and power devices by engineering defect behavior under varying alloy compositions and processing conditions.

\section*{Supplementary Materials}
The Supplementary Material provides a detailed description of the equation of state (EOS) calculations and MCMD simulations for the pure nitride systems, along with comparisons of the computed values to those reported in the literature. In addition, the Supplementary Material includes structural representations of the lowest- and highest-energy nitrogen Frenkel pair configurations for the Al$_{0.25}$Ga$_{0.75}$N alloy.

\begin{acknowledgments}
This work was supported by the Air Force Office of Scientific Research under Award No. FA9550-22-1-0308. 
\end{acknowledgments}

\clearpage
\bibliography{references}

\end{document}